# Decoupling bimolecular recombination mechanisms in Perovskite thin films using photoluminescence quantum yield

Robert Lee Chin[1], Arman Mahboubi Soufiani[1], Paul Fassl[2], Jianghui Zheng[1,3], Eunyoung Choi[1], Anita Ho-Baillie[1,3], Ulrich Paetzold[2], Thorsten Trupke[1], Ziv Hameiri[1]

[1]University of New South Wales, Sydney, Australia
[2]Karlsruher Institut für Technologie, Karlsruhe, Germany
[3]The University of Sydney, Sydney, Australia

## Abstract

We present a novel analytical model for analysing the spectral photoluminescence quantum yield of non-planar semiconductor thin films. This model considers the escape probability of luminescence and is applied to triple-cation perovskite thin films with a 1-Sun photoluminescence quantum yield approaching 25%. By using our model, we can decouple the internal radiative, external radiative, and non-radiative bi-molecular recombination coefficients. Unlike other techniques that measure these coefficients separately, our proposed method circumvents experimental uncertainties by avoiding the need for multiple photoluminescence measurement techniques. We validate our model by comparing the extracted implied open-circuit voltage, effective luminescence escape probabilities, absorptivity, and absorption coefficient with values obtained using established methods and found that our results are consistent with previous findings. Next, we compare the implied 1-Sun radiative open-circuit voltage and radiative recombination current obtained from our method with literature values. We then convert the implied open-circuit voltage and implied radiative open-circuit voltage to the injection-dependent apparent-effective and apparent-radiative carrier lifetimes, which allow us to decouple the different recombination coefficients. Using this lifetime analysis, we predict the efficiency losses due to each recombination mechanism. Our proposed analytical model provides a reliable method for analysing the spectral photoluminescence quantum yield of semiconductor thin films, which will facilitate further research into the photovoltaic properties of these materials.

## 1 Introduction

The detailed balance limit represents the realistic efficiency limit of a single-junction solar cell, wherein all losses of photo-generated charge carriers are due to band-to-band radiative and Auger recombination in the absorber/active layer [1]. Driving the performance closer to the detailed-balance limit requires eliminating extrinsic recombination, caused by defects in the photovoltaic absorbers and at their interfaces [2]. Hence, the ability to accurately quantify radiative and non-radiative recombination is highly relevant for optoelectronic devices. However, there are many nuances involved in accurately characterising the recombination [3]–[5]. Perovskite thin films (PTF) are an exemplary material to study for this purpose.

In PTFs, the recombination losses are commonly determined from measurements of the emitted photoluminescence (PL) [3], [5]. The fundamental recombination metrics are the excess-carrier ($\Delta n$) decay, often determined from the time-resolved PL (TR-PL) decay [6], the implied open-circuit voltage ($iV_{OC}$), and the radiative implied open-circuit voltage ($iV_{OC,rad}$), frequently determined from the Suns-PL quantum-yield (Suns-PLQY) [2], [7], a non-invasive, non-destructive method. Ideally, $\Delta n$ decay measures the total recombination rate and can be analysed to extract the recombination coefficients, which quantify the strength of selected recombination processes [6]. As the implied voltages are a measure of the implied device performance, they, in fact, depend on the recombination coefficients. Thus, for a deep understanding of the implied device performance, one first needs to determine the recombination coefficients, particularly the radiative recombination coefficient.

The radiative recombination rate, $R_{rad}$, is parameterised by the *internal radiative* recombination coefficient, $B_{rad,int}$, also called the internal radiative bimolecular or *B*-coefficient [1], [6]. $R_{rad}$ and $B_{rad,int}$ are linked by the equation $R_{rad} = B_{rad,int} \cdot (np - n_i^2)$, where $n$ ($p$) is the electron (hole) density, $n_i$ is the thermal equilibrium carrier density, and $(np - n_i^2)$ is the net or excess electron-hole density product. The "bimolecular" descriptor indicates that two free carriers recombine. $B_{rad,int}$ can be calculated directly from the photon energy ($\hbar\omega$) dependent absorption coefficient, $\alpha(\hbar\omega)$, via the van Roosbroeck-Shockley (vRS) equation [8], which equates the thermal equilibrium volume-rates of band-to-band photon absorption and spontaneous emission. For the most well-studied PTF composition, MAPbI$_3$, $B_{rad,int}$ is estimated in the order of $10^{-9}$ cm$^{-3}$.s$^{-1}$ [1], [6]. However, for PTFs such estimations neglect several important effects. Simbula *et al.* recently demonstrated that the exciton absorption peak and polaronic effects should be included in the vRS equation [9]. Several studies on PTFs have also shown that most PL/absorption measurements are affected by photon scattering artifacts, which redshifts the absorption band-edge [10]–[12]. Thus, published values for $B_{rad,int}$ calculated from the vRS equation need to be re-evaluated.

Furthermore, the radiative recombination rate reflected in luminescence measurements ($R_{rad,ext}$) is parameterised by the *external radiative B*-coefficient, $B_{rad,ext}$, which is affected by photon reabsorption: $R_{rad,ext} = B_{rad,ext} \cdot (np - n_i^2)$. $B_{rad,ext}$ needs to be used because only a fraction of the spontaneous emission photons eventually escapes the sample as PL, whilst the remainder is recycled [6], [11], [13], [14]. For thin semiconductor films, such as perovskites, with no parasitic absorbing layers, $B_{rad,ext}$ is equal to $B_{rad,int}$ multiplied by an effective escape probability, $\overline{p}_e$ [15]. Thus, $B_{\text{rad,ext}} = \overline{p}_e \cdot B_{\text{rad,int}}$, noting that $B_{\text{rad,int}}$ is an intrinsic material property that can be experimentally determined from the vRS equation. However, most previous estimates of $\overline{p}_e$ have assumed planar films [6], [16], [17], whereas PTFs are non-planar with a root-mean-square (RMS) surface roughness typically in the range of 20 to 50 nm ($\approx$10% of a 500 nm thick absorber) [18], see Section 2.5 of the Supplemental Material. In that regard, Fassl *et al.* recently demonstrated that $\overline{p}_e$ is three to four times larger than previous estimates and strongly depends on the sample thickness and surface roughness [11].

In PTFs, the situation is further complicated by the observation of non-negligible *non-radiative* bimolecular recombination (NRBR), parameterised by the non-radiative *B*-coefficient, $B_{\text{nr}}$ with corresponding recombination rate equal to $R_{\text{Bnr}} = B_{\text{nr}} \cdot (np - n_i^2)$. Numerous publications, using time-resolved decay measurements including TR-PL [9], [12], [19], [20], time-resolved microwave conductance (TR-MC) [17], and transient absorption (TAS) [12] have demonstrated the existence of a NRBR mechanism. However, these time-resolved techniques all measure the *net* recombination rate. Therefore, the directly accessible *B*-coefficient is the total *B*-coefficient, $B_{\text{tot}} = B_{\text{nr}} + B_{\text{rad,ext}}$. It is also worth mentioning that standard PLQY analysis provides the ratio of the external radiative recombination to the total recombination rate, $\text{PLQY} = \frac{R_{\text{rad,ext}}}{R_{\text{tot}}}$ [19], [21]. Therefore, if bimolecular recombination is dominant, $\text{PLQY} = \frac{R_{\text{rad,ext}}}{R_{\text{rad,ext}} + R_{\text{Bnr}}} = \frac{B_{\text{rad,ext}} \cdot (np - n_i^2)}{B_{\text{rad,ext}} \cdot (np - n_i^2) + B_{\text{nr}} \cdot (np - n_i^2)} = \frac{B_{\text{rad,ext}}}{B_{\text{tot}}}$. If a time-resolved technique and Suns-PLQY are measured on the same sample [12], $B_{\text{rad,ext}}$ can be decoupled, yet $B_{\text{rad,int}} = B_{\text{rad,ext}}/\overline{p}_e$ may be incorrect if $\overline{p}_e$ is calculated assuming a planar film [11].

In this study, we propose an analysis technique based on steady-state Suns-PLQY, which involves measuring the absolute spectral PL, $\phi_{\text{PL}}(\hbar\omega)$, at different steady-state incident photon flux, $\phi_{\text{ex}}$. We also present a new analysis method that decouples the internal radiative, external radiative, and non-radiative *B*-coefficients. This technique integrates the recently developed rigorous escape probability model by Fassl *et al.* [11] and relies on the conversion of implied voltages into carrier lifetimes and, thus, recombination coefficients. It is primarily based on steady-state Suns-PLQY measurements, as opposed to a combination of, for example, Suns-PLQY and TR-PL [21]. This has the advantages of not having to probe the same region multiple times and not changing the excitation conditions between different measurements, which may add systematic uncertainties.

The proposed analysis technique is validated using state-of-the-art triple-cation PTFs, $\text{Cs}_{0.05}\text{FA}_{0.79}\text{MA}_{0.16}\text{Pb}(\text{I}_{0.83}\text{Br}_{0.17})_3$ denoted as Br17. This study aims to answer the following research questions:

1. Can the *B*-coefficients be decoupled using the proposed technique and how do the results compare to published data?
2. What is the effect of the non-radiative recombination, particularly the NRBR, on the implied photovoltaic efficiency parameters?

It is emphasised that the analysis techniques presented in this study should be applicable for other PTF compositions as well as other non-planar semiconductors exhibiting a significant $B_{\text{nr}}$.

## 2 Theory and Analytical Methods

First, we establish the equivalence between Suns-$iV_{OC}$, injection-dependent carrier lifetime, and implied light current-density voltage ($J$-$V$) curves. Next, we introduce the concept of an apparent $\tau_{app}$ ($\Delta n_{app}$) and apparent effective carrier lifetime ($\tau_{app,eff}$), which account for the injection-dependent doping by bulk defects. Using these concepts, we then employ Suns-PLQY measurements to disentangle the external radiative and non-radiative $B$-coefficients.

## 2.1 Suns-$iV_{OC}$, Injection-Dependent Lifetime, and Implied Device Performance

To extract the $B$-coefficients, the key concept of equivalence between Suns-$iV_{OC}$ and the injection-dependent carrier lifetime is used. In this study, Suns-PLQY is converted into Suns-$iV_{OC}$ using Equation (1), where $iV_{OC}$ is related to ($np - n_i^2$) [22]:

$$iV_{OC} = k_B T \cdot \ln\left[\frac{(np - n_i^2)}{n_i^2} + 1\right] \tag{1}$$

$$(np - n_i^2) = \Delta n \cdot \left(\Delta n + N_{dop}\right) \tag{2}$$

$N_{dop}$ is the background doping density from thermally ionised defects. The optical generation rate, $G$, is related to the spectral incident photon flux per energy interval, $\phi_{ex}(\hbar\omega)$, through Equation (3), which is a depth-averaged value of $G$ for an excitation power spectrum with units of W·cm$^{-2}$·eV$^{-1}$ and absorptivity spectrum Abs($\hbar\omega$):

$$G = \frac{1}{W} \int \frac{\text{Abs}(\hbar\omega) \cdot \phi_{ex}(\hbar\omega)}{\hbar\omega} \, d\hbar\omega \tag{3}$$

By considering the generation rate under a reference spectrum, typically the AM1.5G spectrum, $G$ can be converted into an equivalent "Suns". Useful information about the expected device performance without series resistance can then be obtained by plotting $iV_{OC}$ as a function of the Suns, also known as the Suns-$iV_{OC}$. The injection-dependent carrier lifetime curve, which is ubiquitous for crystalline silicon (c-Si) photovoltaics but has been less widely applied to non-c-Si semiconductors, including PTFs, provides a complementary way of representing the implied Suns-$iV_{OC}$. The carrier continuity equation [Equation (4)] relates $\Delta n$ to the recombination rate $R_i$, which can be further related to the carrier lifetime $\tau_i(\Delta n)$ using Equation (5):

$$\frac{\partial \Delta n}{\partial t} = G - \sum_i R_i \tag{4}$$

$$\tau_i(\Delta n) = \frac{\Delta n}{R_i} \tag{5}$$

The injection-dependent lifetime is a useful metric for measuring the absolute recombination rate, unlike other recombination metrics such as PLQY, which measure only the relative recombination rate [4]. It represents the average duration of existence for an excess carrier resulting from a particular recombination mechanism.

Recombination rates are expressed using a pre-factor, $K(\Delta n)$ with units of cm$^3$.s$^{-1}$, multiplied by the net electron-hole product, as shown in Equation (6):

$$R(\Delta n) = K(\Delta n) \cdot \Delta n \big(\Delta n + N_{\text{dop}}\big) \qquad (6)$$

For some recombination mechanisms, $K(\Delta n)$ can be approximated by a constant value over a wide injection range, while for others, such as band-to-band radiative and Auger recombination, $K(\Delta n)$ is a function of $\Delta n$. Regardless of the injection-dependence of $K(\Delta n)$, the net electron-hole product is a quadratic function of $\Delta n$, with $N_{\text{dop}}$ defining the low injection (LI, $\Delta n \ll N_{\text{dop}}$) and high injection (HI, $\Delta n \gg N_{\text{dop}}$) regions. Considering Equation (5), this means that carrier lifetimes are usually injection dependent.

The effective lifetime, which considers the impact of all relevant recombination mechanisms at each injection level, can be experimentally determined using Equation (7):

$$\frac{1}{\tau_{\text{eff}}} = \sum_i \frac{1}{\tau_i} \qquad (7)$$

The subscript $i$ denotes each recombination process. Suns-PLQY, measured under steady-state excitation, can be used to calculate the effective lifetime by setting $\sum R = G$ as the steady-state solution of Equation (4): $\tau_{\text{eff}} = \Delta n / G$. This can be further used to predict the photovoltaic device efficiencies by converting Suns-$iV_{\text{OC}}$ into the device light-$iJ$-$V$ curve using Equations (8) and (9). In Equation (9), $\sum R$ can be replaced by selected recombination rates to quantify efficiency losses due to specific recombination mechanisms, such as setting $\sum_i R_i = R_{\text{rad,ext}}$ to correspond to the radiative limit.

$$iV = iV_{OC} \qquad (8)$$

$$iJ = qW(G_{1\text{Sun}} - G) \qquad (9)$$

**Figure 1** shows the Suns-$iV_{\text{OC}}$, injection-dependent lifetime, and implied efficiency-voltage (*iJ-V*) curves for a hypothetical PTF. The circle markers in the sub-figures are colour-coded to represent the same operating conditions for Suns-$iV_{\text{OC}}$, lifetime-$\Delta n$, and *iη-V* curves. The red triangle indicates the maximum power point (MPP), and the dashed line represents the transition from LI to HI. For this example, the MPP is closer to the LI regime, indicating that $N_{\text{dop}}$ is relevant for the fill-factor (*FF*). The black diamond represents the open-circuit condition, which is close to the HI condition, thus, the $iV_{OC}$ is relatively unaffected by $N_{dop}$. The solid orange line represents the detailed balance limit, indicating the maximum performance attainable when only Auger and radiative recombination (with photon recycling) is present.

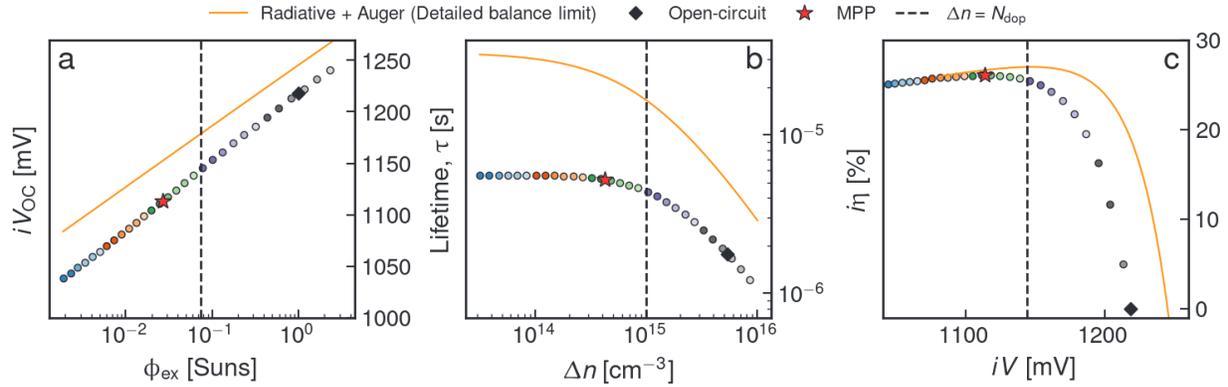

**Figure 1** Simulated curves (simulation parameters in Section 2.6 of the Supplemental Material) with colour-coded markers, showing the 1-Sun and MPP, and $\Delta n = N_{dop}$ conditions for: a) Suns-$iV_{OC}$, b) injection-dependent lifetime, and c) *iη-V*.

## 2.2 Apparent Carrier Density and Apparent Lifetime

In the previous section, we established the equivalence of Suns-$iV_{OC}$ and injection-dependent effective lifetime. In this section, we extend the concept of effective lifetime and excess carrier density to the apparent injection-dependent lifetime ($\tau_{app}$) and apparent excess carrier density ($\Delta n_{app}$) to account for the doping induced by carrier injection when there are bulk defects, explained in the following paragraph.

Equation (2) in the previous section contains the thermal doping density $N_{dop}$ and assumes that the excess electron and excess hole densities are equal ($\Delta n = \Delta p$). However, the doping density measured using dark conductance in PTFs is fairly low, typically in the range $N_{dop} < 10^{13}$ cm$^{-3}$ [23], while the bulk defect densities $N_t$ are significantly larger, estimated to be in the range of $10^{15}$ cm$^{-3}$ to $10^{16}$ cm$^{-3}$ [24]. When excess carriers are generated inside the absorber they can ionise the defects, inducing an injection-dependent bulk doping density ($\Delta n \neq \Delta p$). In previous studies, this effect was called "photo-doping" [21], [25], [26]. In this study, we clarify that this effect is caused by excess carriers generated by any means, not only optical injection, and therefore refer to it more generally as "$\Delta n$-doping", symbolised by $\Delta n_t = \Delta n_t(\Delta n)$. The total doping density $n_t$ is the sum of the excess carrier and thermally induced terms, $n_t = \Delta n_t(\Delta n) + N_{dop}$, resulting in:

$$(np - n_i^2) = \Delta n \cdot \left( \Delta n + \Delta n_t + N_{dop} \right) \qquad (10)$$

The exact injection-dependence of $n_t(\Delta n)$ depends on the defect recombination parameters: electron and hole capture rates ($c_n$ and $c_p$), defect energy level ($E_t$), and $N_t$. Hence, the injection-dependence of $n_t(\Delta n)$ is quite complex even for the simplest defect recombination models such as the bulk Shockley-Read-Hall (SRH) recombination model [27]. Consequently, there is no simple equation relating $iV_{OC}$ to $\Delta n$ when $\Delta n_t$ is significant. To resolve this issue, we define an apparent excess carrier density (lifetime), interpreted as the geometric average of the electron and hole carrier densities (lifetimes):

$$\Delta n_{app} = -\frac{N_{dop}}{2} + \sqrt{\left(\frac{N_{dop}}{2}\right)^2 + (np - n_i^2)} \tag{11}$$

$$\tau_{app} = \frac{\Delta n_{app}}{G} \tag{12}$$

The apparent carrier lifetime The concept of apparent carrier density and lifetime originates from quasi-steady-state and transient photoconductance lifetimes in c-Si photovoltaics, where $\Delta n$-doping causes an artificially long lifetime at low $\Delta n$ due to the recombination of thermally emitted trapped carriers, known as the "trapping artifact" [28]–[30]. Although our Suns-PLQY measurements do not appear to be severely affected by the injection-dependence of the $\Delta n$-doping due to the HI situation, we show in Section 19 of the Supplemental Material that in these PTFs, the $\Delta n$-doping can be relevant at carrier densities below $10^{15}$ cm$^{-3}$.

## 2.3 *B*-Coefficients Determined from PLQY

This section discusses the analysis of the Suns-PLQY of a non-planar semiconductor film to extract the radiative and non-radiative *B*-coefficients. The analysis involves extracting $iV_{OC}$ and $iV_{OC,rad}$ from the PLQY and converting them into $\tau_{eff,app}(\Delta n_{eff,app})$ and $\tau_{app,ext-rad}(\Delta n_{app})$, respectively. The assumptions underlying this analysis are discussed in Section 2.5 of the Supplemental Material. The spectral PL flux emitted from a semiconductor slab into a sphere is represented by the Lasher-Stern-Würfel (LSW) equation [31]:

$$\phi_{PL}(\hbar\omega) = \text{Abs}(\hbar\omega)\,\frac{(\hbar\omega)^2}{2\pi^2\hbar^3 c_0^2}\,\frac{1}{\exp\left[\dfrac{\hbar\omega - iV_{OC}}{k_B T}\right] - 1} \tag{13}$$

The LSW equation can be rewritten as Equation (14) for moderate carrier densities:

$$\phi_{PL}(\hbar\omega) = \text{Abs}(\hbar\omega) \cdot \phi_{BB}(\hbar\omega) \cdot \exp\left[\frac{iV_{OC}}{k_B T}\right] \tag{14}$$

The chosen form of the spectral photon flux from a blackbody, $\phi_{\text{BB}}(\hbar\omega)$, assumes emission into the sphere:

$$\phi_{\text{BB}}(\hbar\omega) \approx \frac{(\hbar\omega)^2}{2\pi^2\hbar^3c_0^2} \cdot \exp\left[-\frac{\hbar\omega}{k_{\text{B}}T}\right] \tag{15}$$

Equation (13) has been used in previous studies to quantify $iV_{\text{OC}}$ by curve-fitting the band-edge region denoted the high-energy region [32]–[34]:

$$\ln\left[\frac{2\pi^2\hbar^3c_0^2}{(\hbar\omega)^2}\frac{\phi_{\text{PL}}(\hbar\omega)}{\text{Abs}(\hbar\omega)}\right] = \frac{iV_{\text{OC}} - \hbar\omega}{k_{\text{B}}T} \tag{16}$$

$\text{Abs}(\hbar\omega)$ can be accurately measured in the high-energy region of the spectrum using standard optical spectroscopic techniques. Then, $iV_{\text{OC,rad}}$ is calculated via the reciprocity relation between the $iV_{\text{OC}}$ and the PLQY [35]:

$$iV_{\text{OC,rad}} = iV_{\text{OC}} - k_{\text{B}}T \cdot \ln[\text{PLQY}] \tag{17}$$

By converting $\phi_{\text{PL}}(\hbar\omega, \phi_{\text{ex}})$ into Suns-$iV_{\text{OC}}$ and Suns-$iV_{\text{OC,rad}}$ curves, the methodology described in Section 2.2 can be used to convert Suns-$iV_{\text{OC}}$ into steady-state $\Delta n_{\text{app}}$ and $\tau_{\text{eff,app}}$, using Equations (1), (11) and (12).

The same procedure is repeated for Suns-$iV_{\text{OC,rad}}$ to extract the apparent external radiative lifetime, $\tau_{\text{app,ext-rad}}(\Delta n_{\text{app}})$. The injection-dependence of $\tau_{\text{app,ext-rad}}$ is given by Equation (18), where $B_{\text{rad,ext}}$ is the inverse of the slope of $\tau_{\text{app,ext-rad}}$ $(1/\Delta n_{\text{app}})$. Note that $\tau_{\text{app,eff}}$ contains information on $B_{\text{tot}}$ as well as any non-bimolecular recombination ($\tau_{\text{app,res}}$), as given by Equation (19). Assuming the expression for the injection-dependence of $\tau_{\text{app,res}}$ is known, $B_{\text{tot}}$ can be extracted from Equation (19) and $B_{\text{non-rad}}$ can then be calculated from the difference $B_{\text{nr}} = B_{\text{tot}} - B_{\text{rad,ext}}$.

$$\tau_{\text{app,ext-rad}} = \frac{1}{B_{\text{rad,ext}} \cdot \Delta n_{\text{app}}} \tag{18}$$

$$\frac{1}{\tau_{\text{app,eff}}} = \frac{1}{B_{\text{tot}} \cdot \Delta n_{\text{app}}} + \frac{1}{\tau_{\text{app,res}}} \tag{19}$$

Another method for quantifying recombination coefficients is the recombination parameter, $J_0$, which is derived by equating the cumulative recombination rate with the diode current density [36]: $J_0 = q \cdot W \cdot n_i^2 \cdot K(\Delta n)$. For bimolecular recombination mechanisms, $K(\Delta n) = B$, and Equation (20) can be used to calculate $J_0$ by substituting $B$ with the desired $B$-coefficient.

$$J_0 = q \cdot W \cdot n_i^2 \cdot B \tag{20}$$

The internal radiative recombination coefficient, $B_{\text{rad,int}}$, can be calculated by two methods. First, from the above $B_{\text{rad,ext}}$ value given $\overline{p}_e$ is known: $B_{\text{rad,int}} = B_{\text{rad,ext}}/\overline{p}_e$. Second, using the vRS equation which relates the which describes the reciprocity between the thermal equilibrium spontaneous emission rate, $R_{\text{sp,0}}(\hbar\omega) = B_{\text{rad,int}} \cdot n_i^2$, and the thermal equilibrium absorption rate. In both methods, there are additional variables to account for. Method 1 requires $\overline{p}_e$, which is not trivial to determine for PTFs which are non-planar. Method 2 requires $n_i^2$ and $\alpha(\hbar\omega)$.

To calculate $n_i^2$ for PTFs, we assume a parabolic band-edge and symmetric electron and hole effective masses, resulting in an analytical expression for $n_i^2$, denoted the "bare" $n_i^2$, which is given in Section 13 of the Supplemental Material. However, for PTFs, polaronic effects or electron-phonon coupling must be accounted for. Recent studies have shown a discrepancy between $B_{\text{rad,int}}$ calculated from the vRS equation and the bare $n_i^2$ due to the neglect of electron-phonon effects. This discrepancy can be resolved by using the polaronic effective mass, $\mu_{\text{pol}}$, instead of the effective mass, $\mu$. This leads to an enhancement of $n_i^2$ relative to the bare approximation by the factor:

$$G_{n_i^2} = \left(\frac{\mu_{\text{pol}}}{\mu}\right)^3 \tag{21}$$

Effective masses in the case of lead halide PTFs are approximated to be symmetric, justified in Section 13.1 of the Supplemental Material. The full expression for $\mu_{\text{pol}}$ can be found in Section 13 of the Supplemental Material. To calculate $n_i^2$, another required knowledge is the semiconductor's electronic bandgap, $E_g$, related to the band-to-band absorption coefficient, $\alpha(\hbar\omega)$. For planar samples, $\alpha(\hbar\omega)$ can be extracted accurately from the absorptivity. However, for non-planar, non-Lambertian samples with significant photon scattering, there is no existing analytical equation relating the absorptivity to $\alpha(\hbar\omega)$. To resolve this issue, we consider the alternative equation for $\phi_{\text{PL}}(\hbar\omega)$ from Fassl *et al.* This equation expresses $\phi_{\text{PL}}(\hbar\omega)$ in terms of the spontaneous emission rate, $R_{\text{sp}}(\hbar\omega)$ [31], [33], and the luminescence escape probability, $P_e(\hbar\omega)$:

$$\phi_{\text{PL}}(\hbar\omega) = a_0 \cdot P_e(\hbar\omega) \cdot R_{\text{sp}}(\hbar\omega) \tag{22}$$

$$R_{\text{sp}}(\hbar\omega, T) = 2 \cdot n_{\text{real}}^2 \cdot \alpha(\hbar\omega) \cdot \phi_{\text{BB}}(\hbar\omega) \cdot \exp\left[\frac{iV_{\text{OC}}}{k_B T}\right] \tag{23}$$

$n_{\text{real}}$ is the real refractive index of the PTF/glass system. Note that the condition of whether surface element emission is into a hemisphere or full sphere should be considered in the solid angle used to determine $\phi_{\text{BB}}(\hbar\omega)$. We can determine the scaling constant $a_0$, by considering that the emitted photon flux is equal to the depth integral of $P_e(\hbar\omega) \cdot R_{\text{sp}}(\hbar\omega)$. Our assumption of uniform carrier densities (see Section 2.5 of the Supplemental Material) across the film thickness means that $a_0$ is equal to $W$. Using the generalised Planck's equation [31], we can define the absolute spectral PL with the Lee-Soufiani-Fassl (LSF) equation, given by:

$$\phi_{\text{PL}}(\hbar\omega) = P_e(\hbar\omega) \cdot R_{\text{sp}}(\hbar\omega) \cdot W \tag{24}$$

where $P_e(\hbar\omega)$ is the probability of a photon escaping from the film. For PTFs, Fassl *et al*. have decomposed $P_e(\hbar\omega)$ into the direct and scattered escape probabilities, $P_{e-d}(\hbar\omega)$ and $P_{e-s}(\hbar\omega)$, respectively [11]. $P_{e-d}(\hbar\omega)$ considers only the escape cones: PTF-to-glass and PTF-to-air, while $P_{e-s}(\hbar\omega)$ represents the escape probability due to photons not directly emitted within the escape cone.

$$P_e(\hbar\omega) = P_{e-d}(\hbar\omega) + P_{e-s}(\hbar\omega) \tag{25}$$

$$P_{e-d}(\hbar\omega) = p_{e-d} \cdot \exp[-\alpha(\hbar\omega) \cdot W_{\text{eff}}] \tag{26}$$

$$P_{e-s}(\hbar\omega) = \frac{(1 - p_{e-d}) \cdot P_s \cdot \exp[-\alpha(\hbar\omega) \cdot z_{\text{avg}}]}{1 - (1 - P_s) \cdot \exp[-\alpha(\hbar\omega) \cdot z_{\text{avg}}]} \tag{27}$$

Equation (26) defines $P_{e-d}(\hbar\omega)$ as the emission probability for PL photons "directly escaping" the film, considering only the escape cones, PTF-to-glass, and PTF-to-air. $p_{e-d}$ is the escape probability considering only the fraction of the escape cone relative to the full sphere, and the reflectance, see section 18 of the Supplemental Material. The exponential term $\exp[-\alpha(\hbar\omega) \cdot W_{\text{eff}}]$ is the attenuation of these escape cone photons propagating across an "effective" film thickness $W_{\text{eff}}$ before escaping. Equation (27) represents the escape probability due to photons that undergo various processes wherein they are internally reflected, reabsorbed (and then may be re-emitted), or scattered out of the film before reabsorption. In Equation (27), the parameters $P_s$ and $z_{\text{avg}}$ are the average fraction of photons and the average respective propagation distance of each scattering event, respectively. By substituting $P_e = 1$ into Equation (24), we can define an "ideal" absolute PL spectrum that is unaffected by photon reabsorption (recycling) and scattering:

$$\phi_{\text{PL,ideal}}(\hbar\omega) = R_{\text{sp}}(\hbar\omega) \cdot W \tag{28}$$

The effective escape probability, $\overline{p}_e$, represents the ratio of the spectrally-integrated PL flux to the spectral- and depth-integrated spontaneous emission rate [15]:

$$\overline{p}_e = \frac{\int \phi_{PL}(\hbar\omega)\,d\hbar\omega}{\int R_{sp}(\hbar\omega) \cdot W\,d\hbar\omega} \tag{29}$$

Using Equations (24) and (28), it can be expressed in terms of the photon-energy-dependent escape probability [11]:

$$\overline{p}_e = \frac{\int P_e(\hbar\omega) \cdot \phi_{PL,ideal}(\hbar\omega)\,d\hbar\omega}{\int \phi_{PL,ideal}(\hbar\omega)\,d\hbar\omega} \tag{30}$$

Similar expressions can be written for effective direct ($\overline{p}_{e-d}$) and effective scattered ($\overline{p}_{e-s}$) escape probabilities by substituting Equation (26) or Equation (27) for $P_e(\hbar\omega)$ in Equation (30), as given in Section 14 of the Supplemental Material.

Fassl *et al.* used spectral PL from confocal microscopy to determine $\alpha(\hbar\omega)$, with reduced lateral scattering compared to conventional techniques. They analytically calculated $\alpha(\hbar\omega)$ by treating the sample as a planar film [11]. However, our confocal measurements of Br17 encountered several issues with this approach: (1) local variations in the bandgap on the micron-scale [11], [37]; (2) an initial systematic light-soaking (LS) procedure before the PLQY measurements induced a redshift of the band-edge relative to the confocal measurements; and (3) residual scattering within the confocal field-of-view (FOV) led to a small (~10 meV) redshift of the absorption edge, which resulted in a significant (~40%) overestimation of $n_i^2$. For this reason, we used a theoretical model for $\alpha(\hbar\omega)$ in our implementation of the LSF equation. The Elliot formulation of the absorption coefficient [38] is the most relevant model for PTFs, accounting for three distinct behaviours of the PTF band-edge absorption [1], [39], [40]: (1) direct-bandgap; (2) excitonic transitions and Coulomb enhancement of the continuum states; and (3) structural and thermal disorder causing spectral broadening. We chose the version of the Elliot formula that uses the hyperbolic secant distribution for the broadening [41], which is given by Equation (31). Equation (32) represents the absorption due to excitonic transitions, while Equation (33) represents the absorption due to continuum states. Here, $\alpha_C(\hbar\omega)$ is influenced by the Coulomb enhancement term, $\xi$, which arises from the existence of excitons, $\alpha_0$ is a scaling parameter, and $E_{ex}$ is the exciton binding energy.

$$\alpha(\hbar\omega) = \alpha_{ex}(\hbar\omega) + \alpha_C(\hbar\omega) \tag{31}$$

$$\alpha_{\text{ex}}(\hbar\omega) = \frac{\alpha_0}{\hbar\omega} \lim_{N \to \infty} \sum_j^N \frac{4\pi\sqrt{E_{\text{ex}}^3}}{j^3} \frac{\text{sech}\left[\dfrac{\hbar\omega - E_{\text{g}} + E_{\text{ex}}/j^2}{E_{\text{u}}}\right]}{\pi E_{\text{u}}} \tag{32}$$

$$\alpha_{\text{C}}(\hbar\omega) = \frac{\alpha_0}{\hbar\omega} \lim_{E' \to \infty} \int_{E_{\text{g}}}^{E'} \underbrace{\frac{2\pi\sqrt{\dfrac{E_{\text{ex}}}{E - E_{\text{g}}}}}{1 - \exp\left[-2\pi\sqrt{\dfrac{E_{\text{ex}}}{E - E_{\text{g}}}}\right]}}_{\xi} \underbrace{\sqrt{E - E_{\text{g}}} \frac{\text{sech}\left[\dfrac{\hbar\omega - E}{E_{\text{u}}}\right]}{\pi E_{\text{u}}}}_{\text{free-carrier}} \, \mathrm{d}E \tag{33}$$

In **Figure 2**, the spectral PL components for a hypothetical PTF is displayed. The total escape probability, $\overline{p}_{\text{e}}$, is represented by the area under the red curve (total spectral PL photon flux) relative to the area under the orange curve (depth-integrated spontaneous emission rate). Additionally, the effective scattered (direct) escape probability, $\overline{p}_{\text{e-s}}$ ($\overline{p}_{\text{e-d}}$), is calculated by comparing the area under the grey (blue) curve (scattered PL flux) and the area under the orange curve.

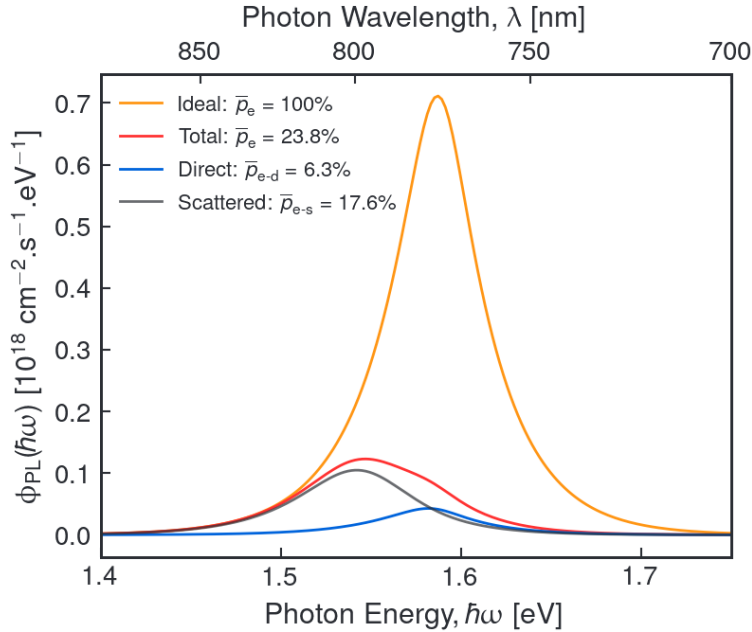

**Figure 2** Simulated absolute spectral PL photon flux for visualising the effective escape probabilities. Simulation parameters are in Section 2.7 of the Supplemental Material. The ideal, total, direct, and scattered absolute spectral PL flux are the orange, red, blue, and grey curves, respectively.

The fitted LSF equation parameters can be used to calculate other important parameters. By equating the LSF and LSW equations, Equation (34) provides a new expression for absorptivity. $J_{0,\text{rad}}$ and $iV_{\text{OC,rad}}$ can be determined from this new absorptivity equation using Equations (35) and (36). The vRS equation, used to calculate $B_{\text{rad,int}}$, is given by Equation (37).

$$\text{Abs}(\hbar\omega) = 2 \cdot P_{\text{e}}(\hbar\omega) \cdot n_{\text{real}}^2 \cdot \alpha(\hbar\omega) \cdot W \tag{34}$$

$$J_{0,\text{rad}} = q \int \text{Abs}(\hbar\omega) \cdot \phi_{\text{BB}}(\hbar\omega) \, \mathrm{d}\hbar\omega \tag{35}$$

$$iV_{\text{OC,rad}} = k_{\text{B}}T \cdot \ln\left[\frac{q \cdot W \cdot G}{J_{0,\text{rad}}} + 1\right] \tag{36}$$

$$\underbrace{\frac{B_{\text{rad,int}} \cdot n_{\text{i}}^2}{\int R_{\text{sp,0}}(\hbar\omega) \, \mathrm{d}\hbar\omega}}_{} = 2 \int n_{\text{real}}^2 \cdot \alpha(\hbar\omega) \cdot \phi_{\text{BB}}(\hbar\omega) \, \mathrm{d}\hbar\omega \tag{37}$$

It is worth highlighting a few of the limitations of this proposed model [11]. First, it is only relevant for samples measured inside a $4\pi$ sr (spherical) integrating sphere. Using a $2\pi$ sr (hemispherical) integrating sphere will discard some of the edge emission from the measurement. Second, the current model only applies for bare films on *non-absorbing* substrates. Third, samples showing more than one material, such as samples with partial phase segregation [42], cannot be analysed with the current model. These limitations are described in more detail in Section 2.5 of the Supplemental Material.

## 3 Results and Discussion

Exemplary results are shown for the intermediate thickness (≈260 nm) PTF, unless stated otherwise. Results for the "thick" and "thin" samples are available Section 4 of the Supplemental Material. First, we validate the LSF equation by comparing the 1-Sun $iV_{\text{OC}}$, effective escape probabilities, photon reabsorption parameters, absorptivity, and $\alpha(\hbar\omega)$ with reference values, which are assumed to be the ground truth. Next, we analyse the injection-dependent lifetime to extract the *B*-coefficients and compare them with published values. We demonstrate the influence of non-radiative recombination, including NRBR, on the implied photovoltaic efficiency. Finally, we provide evidence that defects at the surfaces cause NRBR and suggest using the $J_0$ parameter as an alternative metric to quantify this type of recombination.

## 3.1 Evaluating the Validity of the LSF Equation

In **Figure 3(a)**, the LSW equation is used to curve-fit the 1-Sun $\phi_{\text{PL}}(\hbar\omega)$, see Section 8 of the Supplemental Material. The resulting 1-Sun PLQY is (26.7 ± 1.4)% comparable to the best-reported values in the range of 35% to 50% [11], [16], [17], indicating that these PTFs are close to the state-of-the-art. **Figure 3(b)** displays the reference absorptivity from SE/T spectroscopy and the back-calculated absorptivity using the LSW equation. To correct for the light-induced bandgap shift (see Section 2), the reference absorptivity is redshifted by 78 meV, see the relevant discussion below and Section 10 of the Supplemental Material. This redshift value of was determined iteratively such that the extracted carrier temperature from the LSW equation is consistent with the known ambient temperature of ~298 K. The photon energy fitting region from 1.57 to 1.65 eV in **Figure 3(a)** avoids the "edge" artefact caused by scattered sub-band-gap photons wave-guiding out of the glass edge, denoted as the plateau at photon energies below 1.55 eV [43], and the region with a low signal-to-noise ratio (SNR) above 1.65 eV. The careful choice of the fitting region is reflected by the low uncertainties of ±1 mV and ±0.4 K for $iV_{OC}$ and $T$, respectively.

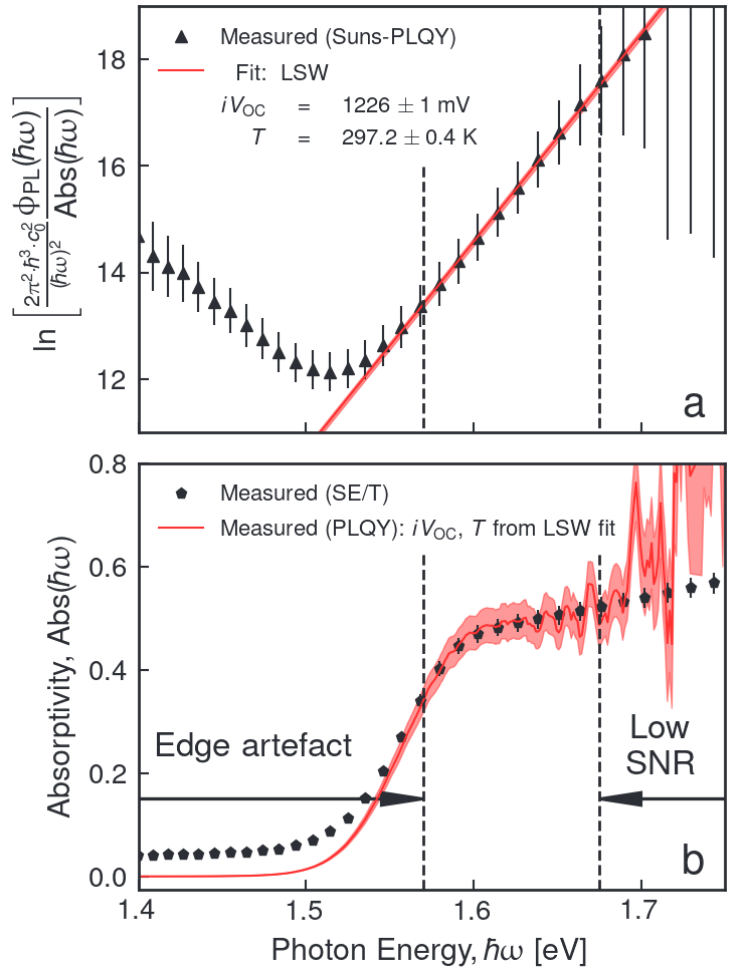

**Figure 3** Intermediate Br17 PTF: a) Curve-fit using Equation (16) from the LSW equation for extracting the 1-Sun $iV_{OC}$ and $T$. The measurement error bars represent the $\pm 3\%$ relative uncertainty in $\phi_{PL}(\hbar\omega)$ and the spectrometer noise floor. For clarity, only every 10$^{th}$ measurement data point is shown. The dashed black lines indicate the photon energy fitting range while the shaded red region represents the total uncertainty of the fit. b) Reference absorptivity

from SE/T and the back-calculated absorptivity using the LSW equation. The SE/T absorptivity is the average of the glass/PTF and air/PTF interfaces, see Section 17 of the Supplemental Material. The shaded red region is the uncertainty in the back-calculated absorptivity.

To perform practical curve-fitting using the LSF equation with the Elliot formula for the absorption coefficient, a total of 12 parameters must be considered, along with an additional band-gap redshift explored in the next section. However, when it comes to practical curve-fitting, some of these parameters must be assigned fixed values. The uncertainties associated with each of these fixed parameters are then incorporated into the overall fit uncertainty. The following dot points outline the values of these fixed parameters and their justifications:

- **Carrier temperature:** Extracted from the carrier temperature from LSW fit vs. $\phi_{\mathrm{ex}}$, $T = 297.9 \pm 0.38$ K, see Section 9 of the Supplemental Material.
- **Scattering fraction:** $P_{\mathrm{s}} = 0.005$, see [11].
- **Real refractive index:** $n_{\mathrm{real}} = 1.7 \pm 0.034$. This value is chosen so the high-energy (> 1.6 eV) absorptivity matches the measured absorptivity. We consider a relative uncertainty of ±2%. Note that this an *effective* refractive index of the PTF/glass system due to waveguide effects in the substrate, see Section 3.6 of the Supplemental Material.
- **Sample thickness:** From SE/T, $W = 262.2 \pm 2$ nm, see Section 1 of the Supplemental Material.
- **Absorption coefficient scaling factor:** From the above band-edge region of $\alpha(\hbar\omega)$ from SE/T, $\alpha_0 = (7.85 \pm 0.16) \times 10^4 \ \mathrm{cm}^{-1} \cdot \mathrm{eV}^{-\frac{1}{2}}$, see the discussion for **Figure 5**.
- **Exciton binding energy:** From the above band-edge region of $\alpha(\hbar\omega)$ from SE/T, $E_{\mathrm{ex}} = 9.0 \pm 0.5$ meV, see **Figure 5** and Section 12 of the Supplemental Material.
- **Effective thickness for the direct emission:** We assumed the bulk diffusion length, $L_{\mathrm{bulk}}$ is much longer than the thickness $W$. Therefore, the average location for the generation of spontaneous emission photons is at $W/2$. We found that a fixed value of $W_{\mathrm{eff}} = 0.3 \cdot W$ gave the most consistent fit quality to the above band-edge absorptivity for the investigated Br17 PTFs. An uncertainty of $\pm 0.25 \cdot W_{\mathrm{eff}}$ was assumed, to account for the surface roughness of several tens of nm, see Section 2.5 of the Supplemental Material. A sensitivity analysis for $W_{\mathrm{eff}}$ is presented in Section 10 of the Supplemental Material.
- **Band-gap redshift:** We found that a redshift, $\Delta\hbar\omega$, of the absorptivity/$\alpha(\hbar\omega)$ photon energy was required for consistent analysis. The redshift value is chosen such that the absorptivity and $\alpha(\hbar\omega)$ from SE/T and the LSF equations are in good agreement and the extracted carrier temperature from the LSW fit is close to the expected ambient temperature of about 25 °C (298.15 K), $\Delta\hbar\omega = -78$ meV. See Section 11 of the Supplemental Material.

Using these fixed parameters, the LSF equation reduces to five free parameters: $iV_{\mathrm{OC}}$, $p_{\mathrm{e-d}}$, $z_{\mathrm{avg}}$, $E_{\mathrm{g}}$, and $E_{\mathrm{u}}$. The curve-fitting results are shown in **Figure 4**. The total fit in red shows good agreement with the measured PL flux and the extracted $iV_{\mathrm{OC}}$ values from the LSF and LSW equations are in excellent agreement: 1224 ± 1 mV, and 1225 ± 1 mV for LSW and LSF, respectively.

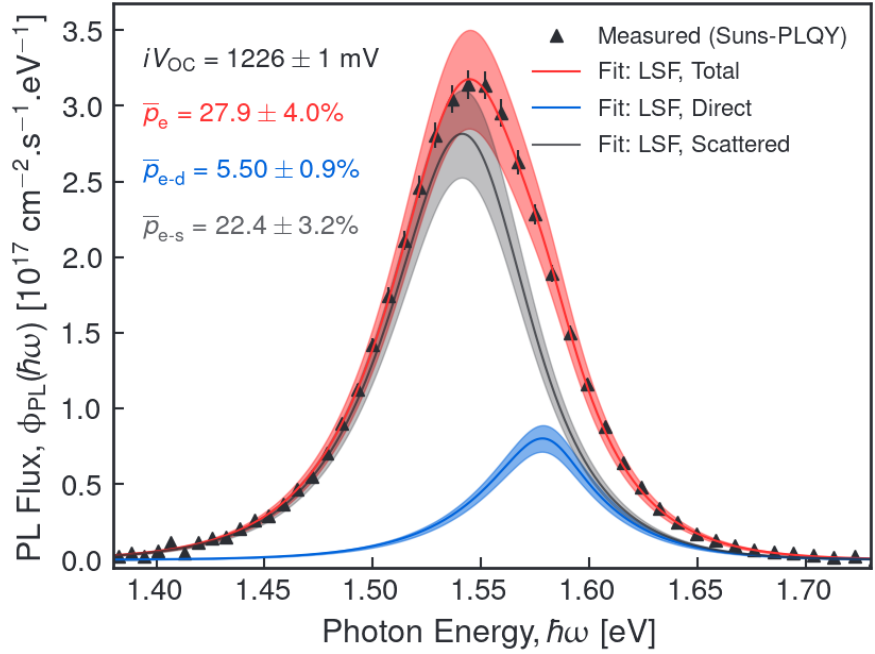

**Figure 4** 1-Sun $\phi_{\mathrm{PL}}(\hbar\omega)$ of intermediate Br17 PTF, curve-fitted using the LSF equation. For clarity, only every 10th measurement data point is shown. Red, blue, and grey curves represent the total, direct, and scattered emission curve-fits, respectively. The shaded regions represent the uncertainties propagated from the uncertainties in the 12 input parameters. The effective escape probabilities, $iV_{\mathrm{OC}}$ and their uncertainties are indicated in the upper left.

A justification for using the Elliot absorption formula with the LSF equation is obtained by comparing the effective escape probabilities and photon reabsorption parameters with reference values published by Fassl *et al*. [11]. These reference values are measured on a MAPbI$_3$ PTF of approximately the same thickness. The reference effective escape probabilities are extracted using the $\alpha(\hbar\omega)$ determined from confocal spectral PL measurements, which are relatively unaffected by scattering artefacts, see Section 5 of the Supplemental Material. Therefore, we interpret similar values for the effective escape probabilities as supporting evidence for the validity of using the Elliot absorption formula in the LSF equation. These values are compared in **Table 1**. Considering the fitting uncertainties, the effective escape probabilities are in good agreement with the reference values. The escape cone probability $p_{\mathrm{e-d}}$ is within the expected range of 4% to 10%, see Section 18 of the Supplemental Material. Only $W_{\mathrm{eff}}$ differs significantly from its reference value by about 43% relative. However, since $W_{\mathrm{eff}}$ and $p_{\mathrm{e-d}}$ are used in the calculation of $\overline{p}_{\mathrm{e-d}}$, yet $\overline{p}_{\mathrm{e-d}}$ agrees with the reference value, this indicates that the curve-fitting is relatively insensitive to the value of $W_{\mathrm{eff}}$. $z_{\mathrm{avg}}$ shows a 25% relative difference which can be accounted for by slightly different surface morphologies, expected due to non-identical fabrication conditions. Given the overall good agreement between the effective escape probabilities and reabsorption parameters, the use of the Elliot formula to model the absorption coefficient appears to be justified.

**Table 1** Effective escape probabilities and reabsorption parameters from the 1-Sun $\phi_{PL}(\hbar\omega)$ LSF curve fitting compared with reference values from [11].

| Study | $\overline{p}_e$ [%] | $\overline{p}_{e-d}$ [%] | $\overline{p}_{e-s}$ [%] | $p_{e-d}$ [%] | $z_{avg}$ [nm] | $W$ [nm] | $W_{eff}$ [nm] |
|---|---|---|---|---|---|---|---|
| This study | 27.2 ± 3.80 | 5.4 ± 0.9 | 21.8 ± 3.1 | 6.4 ± 0.2 | 26.3 ± 0.5 | 262.2 ± 2.0 | 104.9 ± 26.2 |
| Fassl *et al.* [11] | 25.4 ± 0.90 | 6.0 ± 0.5 | 19.4 ± 0.8 | - | 35.8 ± 2.0 | 260 | 241.7 ± 9.7 |

**Figure 5(a)** shows the spectral absorptivity from the LSF curve fit and Equation (34) denoted as the LSF absorptivity, compared with reference measurements from SE/T. The LSF absorptivity is divided into direct and scattered components from Equations (26) and (27), respectively. Below the band-edge, the edge artefact in the SE/T absorptivity prevents a direct comparison. However, this edge artefact can be corrected to first order assuming an offset of the form $a \cdot \hbar\omega + b$, where $a$ and $b$ are constants. This first order corrected SE/T absorptivity (grey stars) agrees well with the LSF absorptivity. Thus, the LSF absorptivity appears to be consistent with the reference absorptivity.

Interestingly, the absorption onset is clearly dominated by the scattered component of the absorptivity. This absorption edge corresponds to spontaneous emission photons which undergo many scattering events and causes the absorption onset to redshift by about 40 meV, as compared to the direct component representing emission from a planar surface. The above band-edge behaviour is contributed by similar fractions of the scattered and direct components, suggesting that the non-planar surface morphology plays an important role in light absorption.

**Figure 5(b)** shows $\alpha(\hbar\omega)$ from the LSF curve-fitting, compared with the SE/T measurement. The $\alpha(\hbar\omega)$ from SE/T is determined by simultaneously curve-fitting SE measurements alongside spectral transmission measurements, see Section 2.3 of the Supplemental Material. In addition to the edge artefact, $\alpha(\hbar\omega)$ from the SE/T measurement appears to be affected by photon scattering of weakly-absorbed photons near the band-edge. To demonstrate this, we calculate the $\alpha(\hbar\omega)$ according to the Beer-Lambert law using the LSF absorptivity [35]: $\alpha_{BL}(\hbar\omega) = -\frac{1}{W} \cdot \ln[1 - \text{Abs}(\hbar\omega)]$. Since the Beer-Lambert law assumes planar surfaces (no photon scattering), we expect $\alpha_{BL}(\hbar\omega)$ matches the SE/T $\alpha(\hbar\omega)$, which also neglects photon scattering below the band-edge. This (purple curve) is compared with the $\alpha(\hbar\omega)$ from SE/T (grey stars, edge-artefact corrected). The obtained good agreement supports our assertion that the $\alpha(\hbar\omega)$ from SE/T is affected by photon scattering near the band-edge. This scattering results in a redshift artefact of about 25.5 meV relative to the Elliot $\alpha(\hbar\omega)$, indicated in the inset of **Figure 5(b)**.

In contrast, the above band-edge $\alpha(\hbar\omega)$ at photon energies > 1.6 eV does not appear to be influenced by any optical artefacts. This can be explained by the fact that in SE/T measurements, the incident photons in this energy range are strongly absorbed by the film, minimizing scattering artefacts. Hence, we manually curve-fitted this region by optimising the combination of $\alpha_0$ and $E_{ex}$. We note that the fixed $E_{ex}$ of 9.0 ± 0.5 meV is in excellent agreement with published values, see Section 12 of the Supplemental Material.

It is worth mentioning that several publications have used an alternative model for $\alpha(\hbar\omega)$ which is the convolution of a stretched exponential function (sub-band-gap absorption) with the ideal density-of-states (DOS), often denoted the Katahara absorption coefficient model [16], [37], [44], [45]. We found this model is not appropriate for PTFs, since it neglects excitonic effects relevant near the band-edge. We provide a comparison of the Katahara absorption coefficient model and the Elliott formula using the LSF equation in Section 7 of the Supplemental Material.

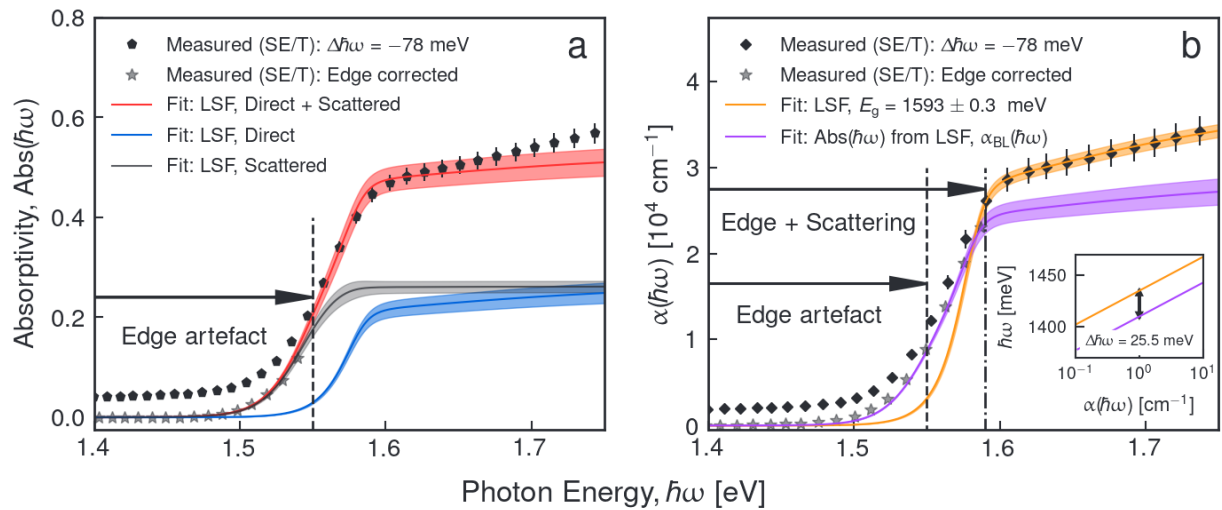

**Figure 5** Comparison of optical parameters from LSF fitting compared with reference measurements from SE/T for intermediate Br17 PTF. Reference measurements are redshifted by 78 meV to correct for the LS-induced band-gap redshift, see Section 11 of the Supplemental Material. a) Absorptivity: The arrow indicates the energy range affected by the edge artefact. b) $\alpha(\hbar\omega)$: Inset shows the redshift of $\alpha_{BL}(\hbar\omega)$, relative to the Elliot $\alpha(\hbar\omega)$. The upper and lower arrows indicate the energy ranges affected by the edge artefact and photon scattering, respectively.

**Figures 6(a)** and **6(b)** present the Suns-$iV_{OC}$ curves obtained from LSW and Rau's equations, and LSF equation, respectively. The two methods show good agreement, with the gradient of the Suns-$iV_{OC}$ curve matching that of the Suns-$iV_{OC,rad}$ curve above 0.3 Suns, indicating similar Ideality factor, $n_{id} = \frac{1}{k_B T}\left[\frac{\partial(G)}{\partial(\phi_{PL})}\right]^{-1}$ [46], and thus significant NRBR. **Figure 6(c)** shows the difference between $iV_{OC}$ obtained from the LSF and LSW equations. These are within 3 mV, considering the fitting uncertainties. **Figure 6(d)** shows the difference between $iV_{OC,rad}$ obtained from the LSF equation compared to the LSW and Rau's equations. These are larger than for $iV_{OC}$ due to the uncertainty propagation of Rau's equation, yet still within 6 mV. Hence, the $iV_{OC}$ and $iV_{OC,rad}$ obtained from the LSF equation are in excellent agreement with the LSW equation (ground truth). Based on the overall excellent agreement between the reabsorption parameters, optical parameters, and Suns-$iV_{OC}$ curves, we determine the LSF equation is a valid model for these PTFs.

Given the validity of the LSF equation, there are a few important consequences. First, non-planar surfaces result in a significant redshift of the absorption edge of ~40 meV. Second, since current models for extracting $\alpha(\hbar\omega)$ from optical measurements erroneously assume planar surfaces, many published values for $E_g$ are likely underestimated by several tens of meV and should be reassessed considering the LSF equation. Third, the full Elliot formula is congruent with both $\phi_{PL}(\hbar\omega)$ and $\alpha(\hbar\omega)$, meaning that the exciton absorption needs to be included when calculating the internal radiative $B$-coefficient via the vRS equation, in contrast to previous studies [1], see Section 15.1 of the Supplemental Material.

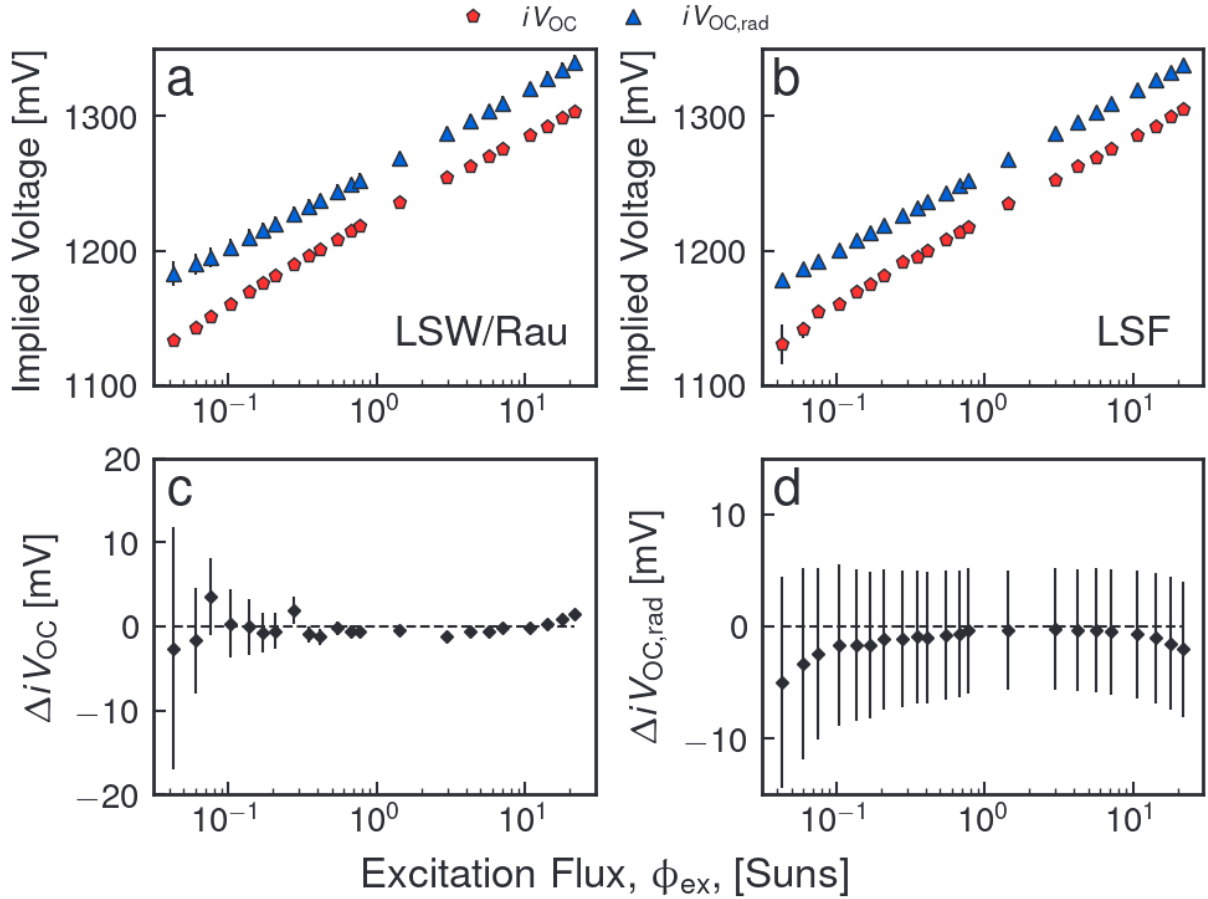

**Figure 6** Suns-$iV_{OC}$ curves for intermediate Br17 PTF, extracted via the a) LSW and b) LSF methods. For this measurement, 1 Sun is equivalent to a current density of 18.7 ± 0.7 mA.cm$^{-2}$. c) Difference in $iV_{OC}$ from LSF compared to LSW equations. d) Difference in $iV_{OC,rad}$ from LSF compared to LSW plus Rau's equations.

## 3.2 Comparison of $iV_{OC,rad}^{1-Sun}$ and $J_{0,rad}$ with Published Values − Resolving Inconsistencies

**Table 2** shows that the values for $iV_{OC,rad}^{1-Sun}$ and $J_{0,rad}$ obtained in this study are 50 mV to 70 mV higher and about ten times higher, respectively, than those reported in previous studies of Br17 [2], [7], [37]. We propose that these significant discrepancies can be explained by the fact that the samples used in this study underwent a LS procedure before measurement, which we believe causes a band-gap redshift. To confirm this assertion, we refer to published data from Stolterfoht *et al.* [2] and Frohna *et al.* [37], who also observed a band-gap redshift. Stolterfoht *et al.* calculated $iV_{OC,rad}$ using the external quantum efficiency [35] of a solar cell instead of the absorptivity [35], assuming perfect carrier collection, while Frohna *et al.* used the reciprocity relationship between luminescence and PLQY, coupled with the $iV_{OC}$ extracted from a full spectrum fitting of $\phi_{PL}(\hbar\omega)$. It is worth noting that Stolterfoht *et al.*'s sample had an estimated thickness of (475 ± 25) nm (assumed 5% relative uncertainty), while for Frohna *et al.*'s the thickness is ~500 nm. As our thick PTF sample had a comparable thickness of (469 ± 4 nm), a direct comparison to these studies is justified.

**Table 2** Comparison of $iV_{OC,rad}^{1-Sun}$, $J_{0,rad}$, and the associated parameters from the LSF equation with published values for Br17 at 300 K. $J_{0,rad}$ values are specified in yocto-Amperes (yA) per cm². Suns-PLQY from Stolterfoht *et al.* [2] corresponds to about 0.88 Suns. Curly brackets indicate the equation(s) used for calculating each quantity.

| Study | $W$ [nm] | $iV_{OC,rad}^{1-Sun}$ [mV] | $iV_{OC}^{1-Sun}$ [mV] | PLQY [%] | $J_{0,rad}$ [yocto-A.cm$^{-2}$] | $J_L$ [mA.cm$^{-2}$] | $E_g$ [meV] |
|---|---|---|---|---|---|---|---|
| This study | 469 ± 4 | 1255 ± 1.9 | 1220 ± 0.5 | 25.9 ± 0.9 | 13.3 ± 1.0 | 20.8 ± 0.8 | 1599 ± 0.3 |
| Frohna *et al.* [37] | 500 | 1310 {17} | 1230 {13} | 4.5 | 1.42 {34, 35} | 21.24 {34, 63} | 1661 ± 10. |
| Stolterfoht *et al.* [2] | 475 | 1317 ± 4.5 {36} | 1187 ± 2.2 {17} | 0.67 ± 0.11 | 1.56 ± 0.25 {34, 35} | 20.63 {34, 63} | 1652 ± 1.2 |

The comparison requires first the appropriate $E_g$ values, corrected from scattering artefacts, for the published results. We reanalysed the 1-Sun $\phi_{PL}(\hbar\omega)$ from Stolterfoht *et al.* [2] using the LSF equation to obtain new values for all the parameters, including an $E_g$ of (1652 ± 1.2) meV. Frohna *et al.* used micro-scale hyperspectral imaging to estimate the bandgap from the absorptivity inflection point, $E_{inflection}$ [47]. We used the $E_g$ ($E_{inflection}$) mapping from the Supplemental Material of [37] to obtain $E_{inflection}$ of (1625 ± 10) meV, where the ±10 meV uncertainty accounts for the μm-scale band-gap disorder. To calculate the true $E_g$, we use the LSF fit parameters for our thick Br17 PTF and Equation (34) to determine the correlation between $E_{inflection}$ and $E_g$. This gives the empirical relationship $E_g = (1.011 \cdot E_{inflection} + 17.87)$ [meV], resulting in the revised $E_g$ value of (1611 ± 10) meV for the data from Frohna *et al.*

Using the photon reabsorption and Elliot formula parameters from the thick PTF LSF curve fitting, $J_{0,rad}$ and $iV_{OC,rad}^{1-Sun}$ are predicted and compared with the values from Stolterfoht *et al.* and Frohna *et al.*, shown in **Figure 7**. Considering the uncertainties, the published values agree with the predicted values. Hence, the observed discrepancies between $J_{0,rad}$ and $iV_{OC,rad}^{1-Sun}$ can be entirely accounted for by an LS-induced band-gap redshift. This justifies the redshift of the SE/T absorptivity and $\alpha(\hbar\omega)$ for the analysis in Section 3.1 and further demonstrates a practical application of the LSF equation for predicting radiative-limit parameters based on changes to $E_g$. Additional evidence that this effect is solely due to a LS-induced change of $E_g$ is presented in Section 11 of the Supplemental Material.

The origin of the band-gap red shift may be related to compression of the lattice structure [48]. One suggestion is charged ions/defects interacting with the lattice anions/cations, leading to lattice compression and, thus, a band-gap red shift.

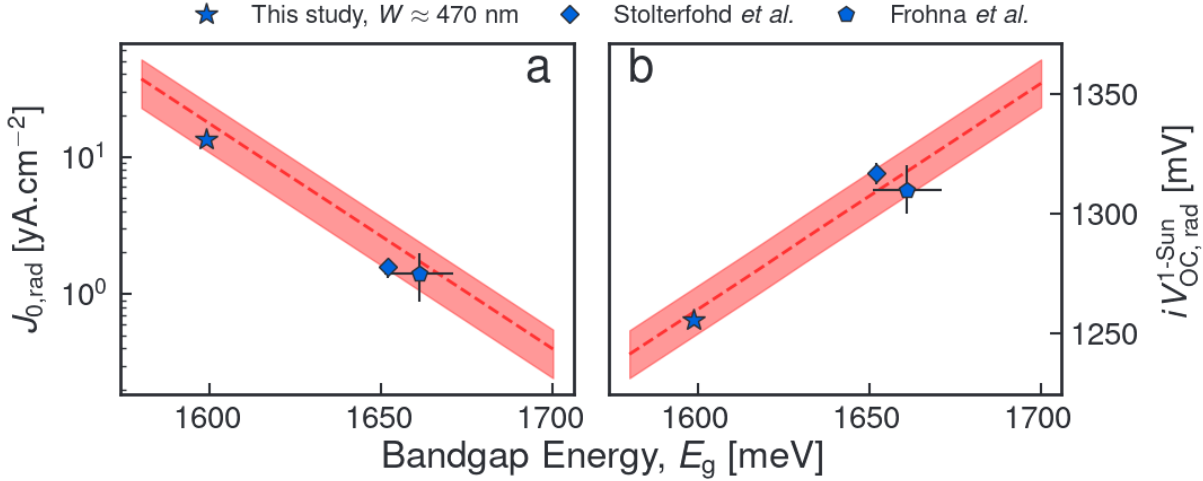

**Figure 7** Predicted $E_g$ dependencies using the LSF equation with $W$ of 475 nm, shaded regions reflect the uncertainty of ±25 nm. a) $J_{0,rad}$ and b) $iV_{OC,rad}^{1-Sun}$.

## 3.3 Injection-Dependent Lifetime Analysis and Implications for the Implied Device Performance

We determine the recombination coefficients by analysing the injection-dependent apparent lifetime curves. We then compare escape probabilities and *B*-coefficients extracted from the injection-dependent lifetime analysis with published values for these parameters. Finally, we evaluate the losses in the implied efficiency parameters relative to the detailed-balance limit.

**Figure 8** shows $\tau_{app,eff}(\Delta n_{app})$ and $\tau_{app,rad}(\Delta n_{app})$, calculated by substituting $iV_{OC}$ and $iV_{OC,rad}$ from the LSF and associated equations into Equations (11) and (12), respectively. $N_{dop}$ for Br17 is in the order of $10^{12}$ cm$^{-3}$ [23], whereas we probe $\Delta n > 10^{14}$ cm$^{-3}$, therefore, we set $N_{dop}$ to zero. Based on the polaronic enhancement of $n_i^2$ from Equations (21) and (64), we calculated $n_i^2 = (13.73 \pm 2.25) \times 10^{10}$ cm$^{-6}$ and $G_{n_{i2}} = 17.48 \pm 2.86$ . Regarding the carrier temperature, injection-dependent lifetimes are typically specified at a constant $T$, since carrier lifetimes are generally injection- *and* temperature-dependent [4]. We assumed the carrier temperature is equal to 300 K, which is about 2 K higher than the 1-Sun temperature, see Section 9 of the Supplemental Material. To curve-fit $\tau_{app,eff}(\Delta n_{app})$, we assume $\tau_{app,res}$ is contributed by: 1) A bulk SRH-type defect [27] with an energy level close to the mid-gap (see Section 19 of the Supplemental Material), with carrier lifetime of $\tau_{SRH}$. Within the range of measured $\Delta n_{app}$, $\tau_{SRH,bulk}$ can be assumed to be injection-independent due to HI. Simulations to justify this assumption are in Section 19 of the Supplemental Material. 2) Band-to-band Auger recombination, with HI apparent Auger lifetime given by $\tau_{Auger,app} = \left(2C_{Auger} \cdot \Delta n_{app}^2\right)^{-1}$. Overall, the apparent effective lifetime can be modelled as:

$$\frac{1}{\tau_{app,eff}} = \tau_{SRH,bulk}^{-1} + B_{tot} \cdot \Delta n_{app} + 2C_{Auger} \cdot \Delta n_{app}^2 \tag{38}$$

This is equivalent to the well-known ABC model [12], [49] with $A = \tau_{\text{SRH,bulk}}^{-1}$, $B = B_{\text{tot}}$, and $C = 2C_{\text{Auger}}$. For $C_{\text{Auger}}$, we used $C_{\text{Auger}} = 7.3 \times 10^{-29} \text{ cm}^6.\text{s}^{-1}$ from Shen *et al.*, assuming a relative uncertainty of ±15 % [50].

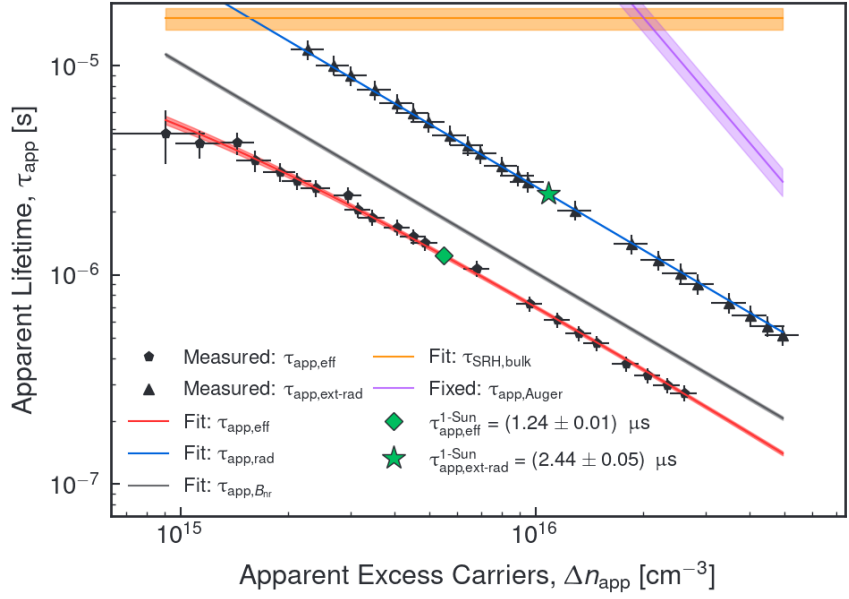

**Figure 8** Injection-dependent apparent lifetimes for intermediate Br17 PTF. $\tau_{B_{\text{nr}}}$ is the apparent non-radiative bimolecular lifetime, $\tau_{\text{app},B_{\text{nr}}} = (B_{\text{nr}} \cdot \Delta n_{\text{app}})^{-1}$. The shaded red and blue regions are the uncertainties in the curve-fitted $\tau_{\text{app,eff}}$ and $\tau_{\text{app,rad,ext}}$, respectively. The green-diamond and green-star markers indicate the 1-Sun $\tau_{\text{app,eff}}$ and $\tau_{\text{app,rad,ext}}$, respectively. $\tau_{\text{app}}$ is between 300 ns and 20 $\mu s$, which justifies the assumption of spatially uniform charge carrier densities discussed in Section 2.5 of the Supplemental Material.

**Table 3** shows the $B$-coefficients extracted from the injection-dependent lifetime analysis and the effective escape probabilities. $B_{\text{rad,ext}}$ and $B_{\text{nr}}$ both *increase* with *decreasing* film thickness. For $B_{\text{rad,ext}}$, this can be explained by considering the effective escape probability, which Fassl *et al.* showed increases with decreasing thickness [11]. For $B_{\text{nr}}$ an *increasing* $B$-coefficient with *decreasing* thickness may be indicative of surface recombination.

**Table 3** $B$-coefficients and effective escape probability from injection-dependent apparent lifetime analysis for all PTF thicknesses. $B_{\text{rad,int}}$ and $B_{\text{nr}}$ are calculated from $B_{\text{rad,int}} = B_{\text{rad,ext}}/\overline{p}_{\text{e}}$ and $B_{\text{nr}} = B_{\text{tot}} - B_{\text{rad,ext}}$, respectively. $\overline{p}_{\text{e}}$ values and their uncertainties are weighted against $\phi_{\text{ex}}$, which minimises uncertainties introduced by a low SNR of the lower $\phi_{\text{ex}}$ Suns-PLQY measurements.

| Thickness | $B_{\text{rad,ext}}$ [$10^{-11}$ cm$^3$.s$^{-1}$] | $\overline{p}_{\text{e}}$ [%] | $B_{\text{rad,int}}$ [$10^{-10}$ cm$^3$.s$^{-1}$] | $B_{\text{tot}}$ [$10^{-11}$ cm$^3$.s$^{-1}$] | $B_{\text{nr}}$ [$10^{-11}$ cm$^3$.s$^{-1}$] |
|---|---|---|---|---|---|
| Thick | 2.54 ± 0.01 | 24.0 ± 4.1 | 1.06 ± 0.18 | 8.09 ± 0.14 | 5.55 ± 0.14 |
| Intermediate | 3.12 ± 0.01 | 27.6 ± 4.5 | 1.13 ± 0.10 | 11.02 ± 0.16 | 7.90 ± 0.16 |
| Thin | 3.05 ± 0.01 | 30.1 ± 5.7 | 1.01 ± 0.19 | 12.37 ± 0.38 | 9.31 ± 0.38 |

For the comparison of the extracted $B$-coefficients with published values, $\overline{p}_e$, $B_{rad,int}$, and $B_{tot}$ values are most often reported. These values have been calculated using absolute spectral PL/simple geometric optics [6], [16], the vRS equation [8], and from the time-decay of quantities related to the excess carrier density, such as TR-PL [6], [19], [51].

A comparison with published $\overline{p}_e$ values is presented in **Table 4**. We note that these $\overline{p}_e$ values are within 0.5% absolute of the $\overline{p}_e$ values calculated from the injection-dependent lifetime (**Table 3**). In the study by Fassl *et al.*, the escape probabilities were estimated to be 3 to 4 times larger than previous estimations assuming geometric optics and/or planar films [6], [9], [12], [16]. Some of these studies, including Braly *et al.* [16] and Staub *et al.* [6], provided enough information for $\overline{p}_e$ to be re-evaluated using the LSF equation. We find that the re-evaluated escape probabilities are about 4 to 5 times larger than the previous estimates, in line with the predictions of Fassl *et al.* In the case of Staub *et al.*, the equation they used to evaluate $\overline{p}_e$ is equivalent to our expression for $\overline{p}_e$ (Equation (30)]. However, we believe $\alpha(\hbar\omega)$ from Staub *et al* is overestimated near the band-edge [see triangle markers, representing the measured $\alpha(\hbar\omega)$, relative to orange curve, representing the Elliot $\alpha(\hbar\omega)$ from the LSF curve-fit in **Figure 47(b)** of Section 15.2 the Supplemental Material] since they assumed a planar film, thus severely underestimating $\overline{p}_e$. Overall, $\overline{p}_e$ for films ranging in thickness from 100 to 500 nm is 20 to 30%.

**Table 4** $\overline{p}_e$ values from the literature. For the data from Braly *et al.* [16], which was re-evaluated with the LSF equation [1-Sun $\phi_{PL}(\hbar\omega)$ for our measurements]. A relative uncertainty in $W$ of ±5% is assumed.

| Study | Composition | $W$ [nm] | Reported $\overline{p}_e$ [%] | Analysis Method | $\overline{p}_e$ from LSF equation [%] |
|---|---|---|---|---|---|
| This study | Br17 | 469 ± 4 | | | 22.5 ± 3.2 |
| | | 262 ± 2 | | | 27.2 ± 3.8 |
| | | 160 ± 6 | | | 30.6 ± 5.6 |
| Fassl *et al.* [19] | MAPbI$_3$ | 260 | | - | 25.4 ± 0.90 |
| | | 160 | | | 28.0 ± 0.80 |
| | | 80 | | | 31.0 ± 0.90 |
| Braly *et al.* [5] | MAPbI$_3$ | 250 ± 13 | 7.4 | geometric optics | 29.7 ± 7.1 |
| Richter *et al.* [63] | MAPbI$_3$ | 200 | 12.7 | geometric optics | - |
| Simbula *et al.* [74] | MAPbI$_3$ | 100 | 6.28 | Equation (25) of [15] | - |
| Staub *et al.* [78] | MAPbI$_3$ | 311 ± 11 | 5.50 | Equation (25) of [15] | 24.9 ± 5.1 |

**Table 5** compares $B_{rad,int}$ from this study with reported values. We find reported $B_{rad,int}$ values are either near to $10^{-10}$ or closer to $10^{-9}$ cm$^3$.s$^{-1}$. Our $B_{rad,int}$ values are within the former range. This difference is attributed to our inclusion of the polaronic effective mass in $n_i^2$, inclusion of photon scattering in $\overline{p}_e$ and correction of scattering artefacts for $\alpha(\hbar\omega)$. **Table 6** shows $B_{rad,int}$ from this study and other studies re-evaluated using the LSF and vRS equations.

**Table 5** Comparison of $B_{rad,int}$ extracted from different methods, $T = 300\ K$.

| Study | Composition | Method | $B_{\text{rad,int}}$ [$10^{-10}$ cm$^3$.s$^{-1}$] | Notes |
|---|---|---|---|---|
| This study | Br17 | LSF + vRS | 0.85 ± 0.15 | Accounts for excitons [$\alpha(\hbar\omega)$], scattering artefacts, polarons ($n_i^2$) |
| Davies *et al.* [1] | MAPbI$_3$ | absorption spectrum + vRS equation | 10.1 | Unscreened, $\alpha(\hbar\omega)$ from PL not using integrating sphere, affected by scattering, no polarons in $n_i^2$ |
| Richter *et al.* [12] | MAPbI$_3$ | TRPL + PLQY + geometric optics | 0.71 | $\overline{p}_e$ assumes planar film |
| Staub *et al.* [6] | MAPbI$_3$ | TR-PL + $\overline{p}_e$ equation [15] | 8.77 ± 0.79 | $\alpha(\hbar\omega)$ affected by scattering, no polarons in $n_i^2$ |
| Zhang *et al.* [52] | MAPbI$_3$ | Hybrid functional + spin-orbit coupling | 0.6 to 1.1 | First principles |

**Table 6** $B_{\text{rad,int}}$ and associated parameters evaluated using the LSF and vRS equations at 300 K. For all studies which did not report the uncertainty in the thickness, we assumed a relative uncertainty of ±5%. The composition MAFA is MA$_{0.07}$FA$_{0.93}$PbI$_3$.

| Study | Composition | $W$ [nm] | $n_{\text{real}}$ | $B_{\text{rad,int}} \cdot n_i^2$ [cm$^{-3}$.s$^{-1}$] | $E_g$ [meV] | $n_i^2$ [$10^{10}$ cm$^{-3}$] | $B_{\text{rad,int}}$ [$10^{-10}$ cm$^3$.s$^{-1}$] |
|---|---|---|---|---|---|---|---|
| This study | Br17 | 469 ± 4 | 1.7 | 7.45 ± 0.57 | 1599 ± 0.3 | 8.86 ± 1.47 | 0.84 ± 0.15 |
| | | 262 ± 2 | 1.7 | 8.53 ± 0.64 | 1594 ± 0.3 | 10.0 ± 1.65 | 0.85 ± 0.15 |
| | | 160 ± 6 | 1.7 | 8.78 ± 0.73 | 1597 ± 0.4 | 10.4 ± 1.81 | 0.84 ± 0.16 |
| Braly *et al.* [5] | MAPbI$_3$ | 250 ± 13 | 1.75 | 2.06 ± 0.27 | 1649 ± 0.8 | 2.0 ± 0.28 | 1.07 ± 0.30 |
| Fassl *et al.* [19] | MAPbI$_3$ | 260 ± 13 | 1.7 | 3.28 ± 0.36 | 1631 ± 0.5 | 3.87 ± 0.53 | 0.42 ± 0.07 |
| Gutierrez *et al.* [26] | MAFA | 780 ± 38 | 1.7 | 29.7 ± 3.78 | 1541 ± 0.7 | 95.8 ± 12.9 | 0.31 ± 0.06 |
| Staub *et al.* [78] | MAPbI$_3$ | 311 ± 11 | 1.7 | 1.42 ± 0.16 | 1659 ± 0.5 | 1.44 ± 0.20 | 0.98 ± 0.17 |
| Stolterfoht *et al.* [82] | Br17 | 475 ± 25 | 1.7 | 1.13 ± 0.1 | 1652 ± 1.2 | 1.81 ± 0.26 | 0.63 ± 0.10 |

Time-resolved measurements are often used to obtain $B_{tot}$, summarised in **Table 14** of the Supplemental Material for compositions with $E_g$ from 1.6 to 1.7 eV. It is possible to estimate $B_{nr}$ using the mean value of $B_{rad,int}$ from **Table 6** $[B_{rad,int} = (0.62 \pm 0.11) \times 10^{-10} \text{ cm}^{-3}.\text{s}^{-1}]$ and $\overline{p}_e$ from **Table 4**, see Section 14 of the Supplemental Material. The $B_{nr}$ evaluated using this approach is shown in **Figure 9**. All the $B_{nr}$ values are positive, spanning $10^{-11}$ to $3 \times 10^{-9}$ cm³.s⁻¹. The $B_{nr}$ values for the Br17 PTFs in this study are moderate, from 0.5 to 1 x $10^{-10}$ cm³.s⁻¹, which is congruent with the excellent PLQY of these samples of up to about 25%. We note that our $B_{nr}$ values increase with decreasing thickness, indicating a surface origin of the NRBR as elucidated in our previous study [53]. In contrast, using the bare $n_i^2$ to calculate $B_{rad,int}$ results in only 43% positive values for $B_{nr}$, indicating that the bare $n_i^2$ is almost certainly underestimated. Hence, it seems that our $B_{nr}$ values are reasonable and the estimated $n_i^2$ is within the appropriate range.

We provide additional evidence for the validity of our $n_i^2$ value in Section 13 of the Supplemental Material. However, we emphasise that accurately and independently determining $n_i^2$, for example, in a similar manner to c-Si with relative uncertainty in $n_i^2$ of less than a few percent [22], [54], warrants a dedicated study.

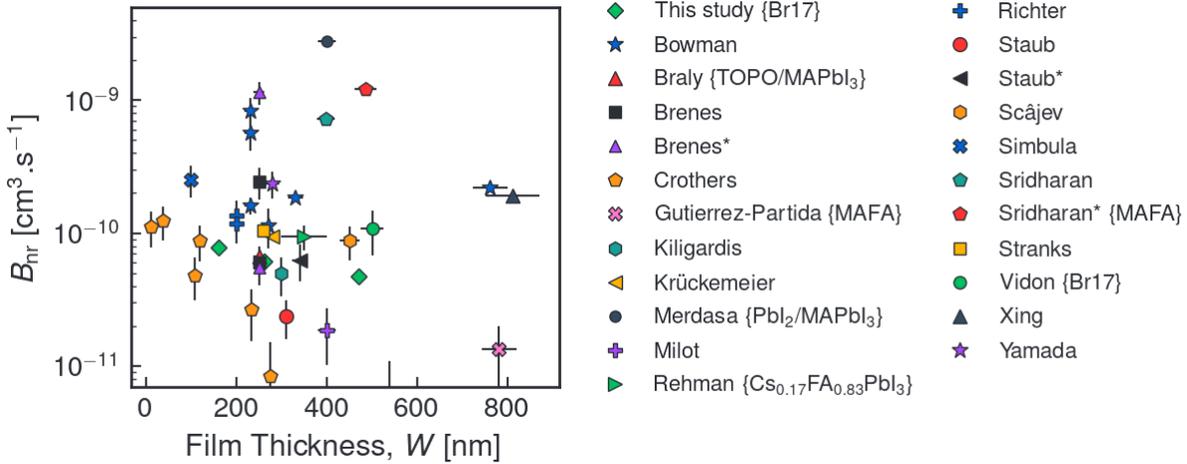

**Figure 9** $B_{non-rad}$ versus film thickness, estimated from published measurements of various PTFs. Curly brackets indicate the composition for non-pure MAPbI₃ samples. We have assumed a relative uncertainty in $B_{tot}$ and the thickness of ±25% and ±5%, respectively, for publications which did not report these uncertainties. Gutierrez-Partida *et al.* [55] and Staub *et al.* [6] also provided $\phi_{PL}(\hbar\omega)$ measurements, therefore, we used the LSF equation to determine $B_{rad,ext}$ in these instances.

**Figure 10** shows the implied efficiency curves, *iη-V*, used to evaluate the *iJ-V* parameter losses due to the non-radiative recombination. These *iη-V* curves are calculated using the injection-dependent apparent lifetime curves following the methodology described in Section 2.1, noting that any systematic uncertainty in $n_i^2$ does not affect the *iJ-V* analysis. The metrics of interest are the $iV_{OC}$ (1-Sun), the implied FF, *i*FF, and the implied MPP efficiency, $i\eta_{MPP}$. We find the bulk SRH recombination leads to only a few mV decrease in $iV_{OC}$ but decreases the *i*FF and $i\eta_{MPP}$ significantly, 0.95% (89.94% to 88.99%) and 0.25% (20.58% to 20.33%), respectively. This can be understood by an increase in $n_{id}$ from $n_{id} \approx 1.05$ ($n_{id} = 1$ for bimolecular recombination) at the $iV_{OC}$ to $n_{id} \approx 1.3$ ($n_{id} = 1.5$ for mid-gap bulk defect, see Section 19 of the Supplemental Material) at the MPP. On the other hand, NRBR causes a significant reduction in $iV_{OC}$ of 31 mV (1257 mV to 1226 mV), leading to a 0.6% (21.18% to 20.58%) absolute reduction in $i\eta_{\mathrm{MPP}}$. Despite the revised values for $B_{\mathrm{rad,int}}$, which are lower than published estimates, we predict that band-to-band Auger recombination has a nearly negligible impact, decreasing $iV_{\mathrm{OC}}$ by only 2 mV and having a negligible effect (< 0.005% absolute change) on $i\eta_{\mathrm{MPP}}$. Hence, for this sample, one need not consider the sum of radiative and band-to-band Auger [56] for calculating the detailed balance limit under the AM1.5G spectrum.

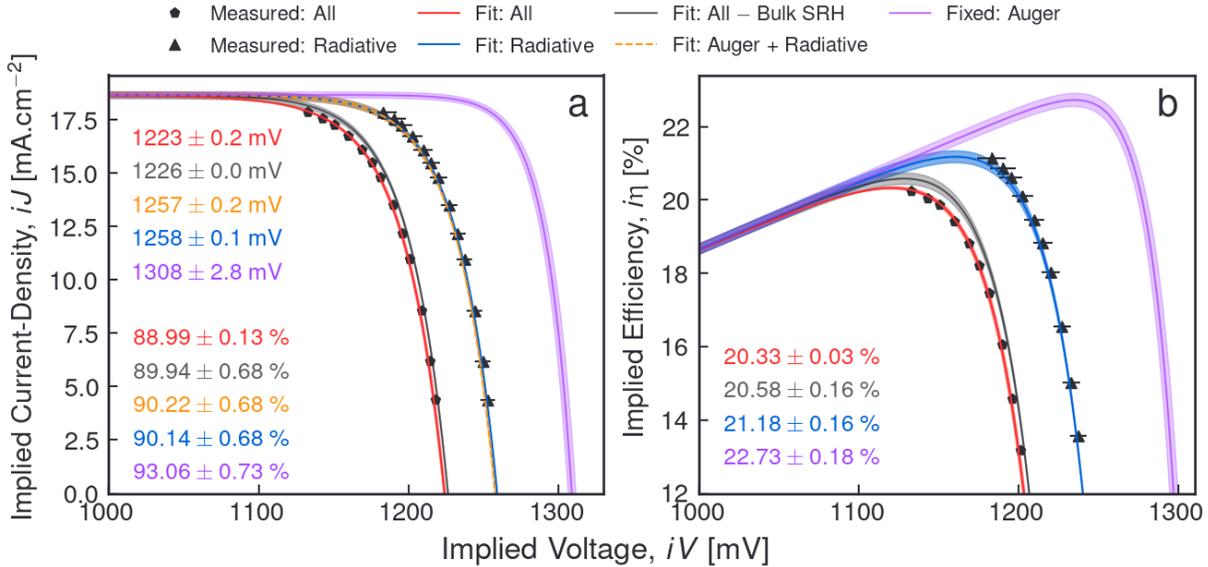

**Figure 10** $iJ$-$V$ parameters for intermediate Br17 PTF. See Section 8 of the Supplemental Material for calculation of $J_{\mathrm{L}}$. a) $iJ$-$V$ curves. Colour-coded $iV_{\mathrm{OC}}^{1-\mathrm{Sun}}$ ($iFF$) values are in the upper (lower) left corner. b) $i\eta$-$V$ curves. For clarity the $i\eta$ curve for the intrinsic recombination is not shown. Colour-coded $i\eta_{\mathrm{MPP}}$ values are indicated in the lower left corner. The radiative + Auger $i\eta$-$V$ curve is not plotted, being nearly identical to the radiative $i\eta$-$V$ (blue) curve.

## 4 Conclusions and Future Works

Our key developments and findings are summarised below. The first three points address our specific research questions from the introduction, whereas the rest of the points are auxiliary developments/findings.

1. The LSF equation was proposed as a new model to analyse the Suns-PLQY of non-planar semiconductor thin films, considering the direct and scattered PL emission. It was validated using state-of-the-art Br17 PTFs by comparison with $iV_{\mathrm{OC}}$ and $iV_{\mathrm{OC,rad}}$ from the LSW and Rau's equations, published data from Fassl *et al.* [11], and the measured $\alpha(\hbar\omega)$ and absorptivity from SE/T. The escape probabilities for typical-thickness PTFs (100 to 500 nm) are estimated from 20% to 30%, which is up to four times larger than published values based on less rigorous estimates.

2. The $B$-coefficients were de-coupled. Published values for $B_{\mathrm{rad,int}}$ appear to be overestimated by an order of magnitude, even without consideration of the exciton absorption which needs to be included in the vRS equation. By accounting for the polaronic enhancement of $n_{\mathrm{i}}^2$ and using the full band-to-band absorption coefficient including the exciton absorption, we calculated a $B_{\text{rad,int}}$ of about $0.85 \times 10^{-10}$ cm$^3$.s$^{-1}$ and an $n_{\text{i}}^2$ enhancement factor of about 17.5. $B_{\text{rad,ext}}$ varies from $2.5 \times 10^{-11}$ to $3 \times 10^{-11}$ cm$^3$.s$^{-1}$, based on extracted $\overline{p}_{\text{e}}$ from the LSF equation from 20% to 30%. $B_{\text{nr}}$ for these samples decreases with increasing PTF thickness and is within the range of 0.5 to $1 \times 10^{-10}$ cm$^3$.s$^{-1}$, which is moderate compared with $10^{-11}$ to $3 \times 10^{-9}$ cm$^3$.s$^{-1}$, estimated from published time-resolved measurements.

3. The injection-dependent lifetime analysis allows accurate quantification of the recombination losses for each sample, relative to the radiative limit. For our Br17 samples, the NRBR leads to a 0.6% absolute reduction in the implied efficiency via a reduction in $iV_{\text{OC}}$ of 30 mV. We were also able to decouple the bulk SRH recombination, which causes up to a 0.24% efficiency reduction due to an 1.1% absolute reduction of the $i$FF ($iV_{\text{OC}}$ reduction of only 3 mV).

4. The concept of an apparent injection-dependent carrier lifetime was established to account for the fact that PTFs have relatively low thermal doping densities ($N_{\text{dop}} < 10^{13}$ cm$^{-3}$) [55], but potentially significant $\Delta n$-doping due to bulk defect densities in the range $N_{\text{t}}$ from $10^{15}$ to $10^{16}$ cm$^{-3}$ [24].

5. Published measurements of $\alpha(\hbar\omega)$ are affected by photon scattering which redshifts the apparent band-edge by 20 to 25 meV. Inserting the Elliot formula into the LSF equation, we model $\alpha(\hbar\omega)$ , which is consistent with both the absolute spectral PL and $\alpha(\hbar\omega)$ measurements from SE/T. The chosen version of the Elliot formula is convolved with the sech-distribution to recover the Urbach-edge behaviour of the below band-edge PL spectra. The Katahara absorption coefficient model used in several previous publications appears to be unsuitable because it cannot model the sharp band-edge caused by exciton absorption.

6. The LSF equation was applied to demonstrate that the initial LS of our Br17 samples resulted in a redshift of the band-gap energy, causing an increase in $J_{0,\text{rad}}$ and concurrent decrease in $iV_{\text{OC,rad}}$ of up to 20 times and 60 mV, respectively. This LS-induced band-gap redshift does not appear to be thoroughly investigated in the literature.

There are many potential future works. Based on the above findings they could include 1) extending the LSF model to include the effect of parasitic absorbing layers, 2) experimentally determining the value of $n_{\text{i}}^2$ for more accurate values for the $B$-coefficients, 3) measuring and comparing $B_{\text{nr}}$ for other PTF compositions/stoichiometries, 4) effect of photodoping on the device performance, and 5) deeper investigation of the effect of the photon scattering on the light absorption for solar cell devices.

## 4.1 Authors' Contributions

- *Robert Lee Chin*: project idea, literature review, Suns-PL measurements, data analysis, draft writing, internal review
- *Arman Mahboubi Soufiani*: project idea, literature review, SE/T + SEM + TEM measurements, arranging sample fabrication, draft writing, internal review
- *Paul Fassl*: Suns-PLQY measurements, technical details regarding the LSF equation, internal review
- *Jianghui Zheng*: sample fabrication, internal review.
- *Eunyoung Cho*: CPD measurements
- *Anita Ho-Baillie*: internal review, project funding
- *Ulrich Paetzold*: technical details regarding the LSF equation, internal review, project funding
- *Thorsten Trupke*: Internal review
- *Ziv Hameiri*: project idea, internal review, project funding

## 4.2 Acknowledgements

The authors thank Dr. Andreas Pusch for constructive discussions throughout the model development process.